\documentclass[fleqn,usenatbib]{mnras}

\usepackage{graphicx}
\usepackage{amssymb}
\usepackage{amsmath}
\usepackage{color}

\usepackage[T1]{fontenc}
\usepackage{ae,aecompl}

\usepackage{newtxtext,newtxmath}

\newcommand{\be}{\begin{equation}}
\newcommand{\ee}{\end{equation}}

\newcommand{\ba}{\begin{eqnarray}}
\newcommand{\ea}{\end{eqnarray}}

\newcommand{\vv}{\upsilonup}

\def\psr{{PSR~B1259--63}}

\def\j06{{HESS~J0632$+$057}}

\title[Radiation-hydrodynamics for HESS~J0632$+$057]{A hydrodynamics-informed, radiation model for HESS~J0632$+$057 from radio to gamma rays}
\author[Barkov and  Bosch-Ramon]{Maxim V. Barkov$^{1,2}$\thanks{Correspondence author: mbarkov@purdue.edu (MVB)} and Valenti~Bosch-Ramon$^{3}$\\
$^{1}$ Department of Physics and Astronomy, Purdue University, West Lafayette, IN 47907-2036, USA\\
$^{2}$ Astrophysical Big Bang Laboratory, RIKEN, 351-0198 Saitama, Japan \\
$^{3}$ Departament de F\'{i}sica Qu\`antica i Astrof\'{i}sica, Institut de Ci\`encies del Cosmos (ICC), \\
Universitat de Barcelona (IEEC-UB), Mart\'{i} i Franqu\`es 1, E08028 Barcelona, Spain
}

\pubyear{2018}

\begin{document}
\date{Received/Accepted}
\maketitle

%%%%%%%%%%%%%%%%%%%%%%%%%%%%%%%%%%%%%%%%%%%%%%%%%%
\begin{abstract} 
Relativistic hydrodynamical simulations of the eccentric gamma-ray binary \j06\, show that the energy of a putative pulsar wind should accumulate in the binary surroundings between periastron and apastron, being released by fast advection close to apastron. To assess whether this could lead to a maximum of the non-thermal emission before apastron, we derive simple prescriptions for the non-thermal energy content, the radiation efficiency, and the impact of energy losses on non-thermal particles, in the simulated hydrodynamical flow. These prescriptions are used to estimate the non-thermal emission in radio, X-rays, GeV, and TeV, from the shocked pulsar wind in a binary system simulated using a simplified 3-dimensional scheme for several orbital cycles.
Lightcurves at different wavelengths are derived, together with synthetic radio images for different orbital phases. The dominant peak in the computed lightcurves is broad and appears close to, but before, apastron. This peak is followed by a quasi-plateau shape, and a minor peak only in gamma rays right after periastron. 
The radio maps show ejection of radio blobs before apastron in the periastron-apastron direction.
The results show that a scenario with a highly eccentric high-mass binary hosting a young pulsar can explain the general phenomenology of \j06: despite its simplicity, the adopted approach yields predictions that are robust at a semi-quantitative level and consistent with multiwavelength observations.
% pointing at the observer 
\end{abstract}
%%%%%%%%%%%%%%%%%%%%%%%%%%%%%%%%%%%%%%%%%%%%%%%%%%


%\maketitle

\section{Introduction}\label{intro}

\j06 is a moderately bright, highly eccentric gamma-ray binary hosting a Be star and a compact object of unknown nature, with an orbital period slightly 
below one year \citep{cas12}. Discovered for the first time as a point-like TeV source of unclear counterpart \citep{aha07}, \j06\, was latter detected in 
radio, X-rays, GeV and TeV energies \citep[e.g.][]{hin09,bon11,2011A&A...533L...7M,ali14,mal16,li17}. The non-thermal emission from \j06\, presents a general 
tendency to peak before apastron, with a broad secondary peak after apastron in X-rays and TeV \citep[e.g.][]{ali14}. The GeV lightcurve is consistent with 
this behaviour \citep{li17}, and  interferometric observations also show a displacement of the radio emission in the sky before apastron 
\citep{2011A&A...533L...7M}.

The fact that the only confirmed gamma-ray (high-mass) binary hosting a pulsar is another eccentric Be binary, \psr\footnote{Very recently, the Be binary 
system PSR~J2032$+$4127, which also hosts a pulsar, has been detected by VERITAS and MAGIC \citep{2017ATel10971....1M}.}, and the similar spectral energy 
distribution between \j06, and \psr\, (the latter when not far from periastron), have been considered indications that \j06\, may be also hosting 
a non-accreting pulsar \citep[see][and references therein]{bbmb17}. If this were the case, the pulsar and the stellar wind would collide, forming a complex 
interaction structure that would be behind the gamma-ray emission from \j06, as it is thought to be the case in \psr\, \citep{aha05}. 

The possibility that \j06 is hosting a pulsar led \cite{bbmb17} to consider, for this source, a scenario similar to that explored by \cite{bb16} for \psr, 
in which an unstable spiral structure, made of shocked flow, forms due to the star and the pulsar wind collision and the effect of orbital motion. 
The authors found that, as in \psr, the pulsar wind energy accumulated between periastron and apastron passage due to spiral formation. Before apastron, 
this energy was quickly released, as the spiral was disrupted by the impact of the pulsar wind in the periastron-apastron direction. The kinematics of 
the disrupted material would be consistent with X-ray observations and the system geometry in the sky \citep{pav15,mil18}. This result led \cite{bbmb17} 
to propose that this behaviour, directly related to the high eccentricity of the system {  ($e\approx 0.83$; see \citealt{cas12})}, may explain the multiwavelength observational properties of \j06. 

In the present work, we study in detail the behaviour of the radiation produced in the shocked pulsar wind along the orbit. For that purpose, we adopt simple 
but physically consistent prescriptions for the non-thermal energy content, the radiation efficiency, and the impact of energy losses on the non-thermal 
particles of the flow. These prescriptions, plus the hydrodynamical information obtained by \cite{bbmb17}, allow us to estimate the amount of radiation 
produced in the shocked pulsar wind in different spectral energies and orbital phases. An alternative scenario, partially based on accretion and focused 
on the GeV emission of \j06, was presented by \cite{yi17}.

The paper is distributed as follows: In Sect.~\ref{hyd}, we briefly present the main features of the hydrodynamical study from \cite{bbmb17}. Then, 
in Sect.~\ref{rad}, we describe the prescriptions adopted, and present the calculation results. Finally, the results are discussed in Sect.~\ref{disc}.

\section{Hydrodynamical approach}\label{hyd}

The computational scheme adopted in \cite{bb16,bbmb17} considers the problem at a distance from the binary much larger than the system semi-major axis ($a$). This allows spherical coordinates, centred at the binary, to reasonably describe the wind interaction on 
those scales. On these scales and geometry, the winds are well reproduced as a conical, radial and relativistic wind (pulsar), surrounded by a radial slow 
wind (star), both supersonic, and centred at the binary location. {  Along the orbit, the pulsar wind cone points in the changing star-pulsar direction. In this way, the cone rotates with time reproducing the orbital motion of period $T=321$~days\footnote{We note that small variations in $T$ \citep[see, e.g.,][]{mor18,mal17} will not significantly alter the results of this work.} with $e=0.83$. If one takes the masses of the Be star and the pulsar equal to $20 M_{\sun}$ and $1.44 M_{\sun}$, respectively, one obtains $a\approx 3.5\times10^{13}$~cm. On scales larger than the binary, the simulation leads to an interaction structure very similar to that found in a 3D simulation when the colliding-wind region within the binary is included \citep{bbp15}}. As noted in \cite{bb16,bbmb17}, the mass of the {  relevant region of the} disc of the Be star is likely much smaller than that of the stellar spherical wind released during one orbit, and thus the dynamics of the former is neglected on the scales of interest. 

Given the large range of spatial scales to cover for eccentric binaries, a fully 3-dimensional (3D) simulation is still too costly even if the binary region 
is simplified in the way explained. One can however assume that the mass, momentum and energy hydrodynamical fluxes get diluted as $1/r^2$ ($r$: distance 
from the binary), and one can then focus on the equatorial plane region of the simulated system (i.e. a very low resolution in $\theta$). This formally yields 
a correct description for that region if most of the material there is supersonic, which is a good approximation at the onset of the spiral and in the apastron 
side of the simulated region, but not everywhere. Nevertheless, the similarity between the results of \citep{bbp15} and those of \cite{bb16,bbmb17} indicates 
that the major features of the hydrodynamics at large scales are already captured by the simplified 3D scheme, which also shortens the computing time by several 
orders of magnitude.

\begin{figure}
\includegraphics[width=0.9\linewidth]{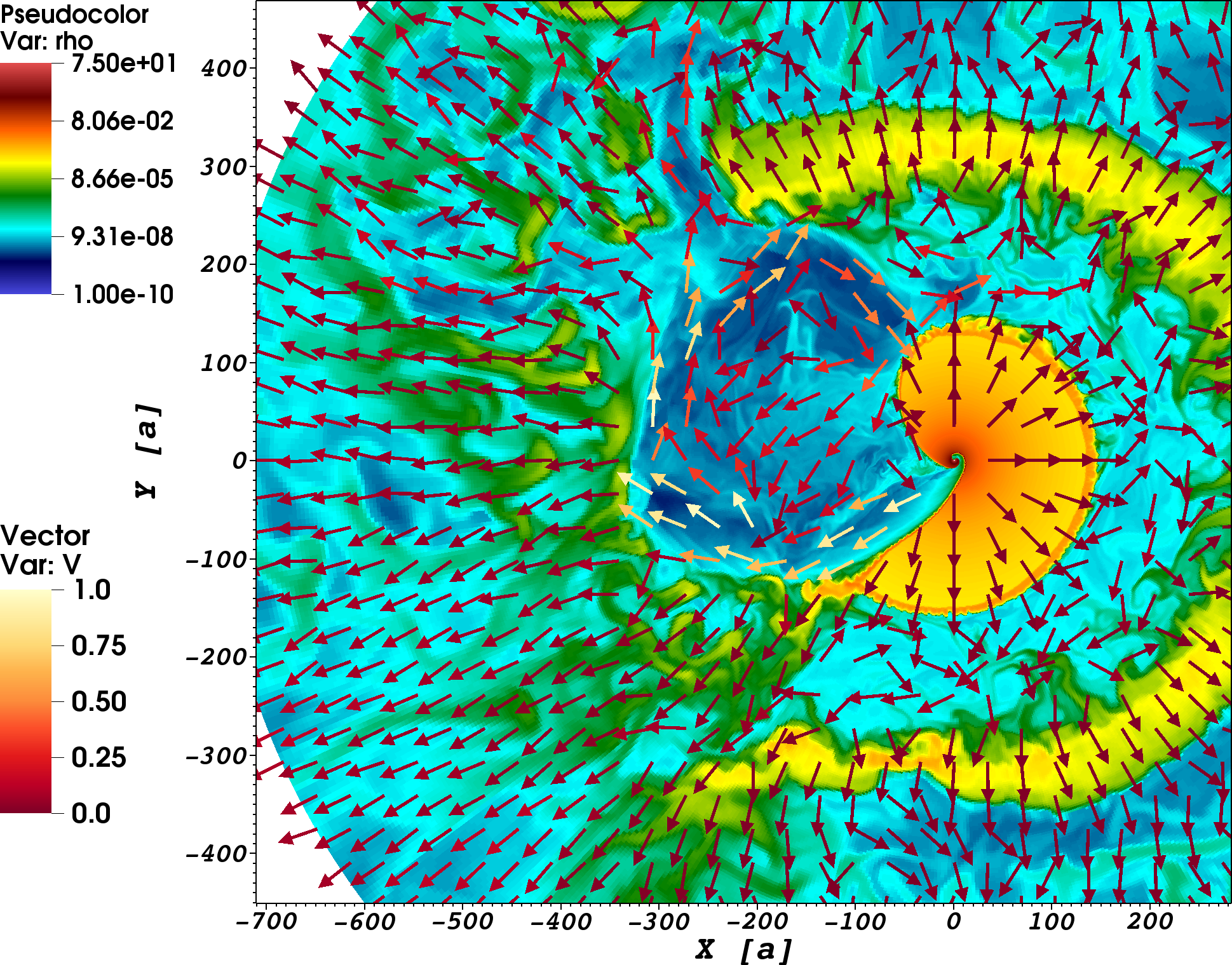}
\caption{Colour map of density, with colored arrows representing the flow direction and three-velocity modulus, for \j06{} at the orbital phase $\phi = 0.026$. 
The apastron side of the grid is to the left. {  The axis units are $a$ ($\approx 3.5\times 10^{13}$~cm).} }
\label{map1}
\end{figure}

\begin{figure}
\includegraphics[width=0.9\linewidth]{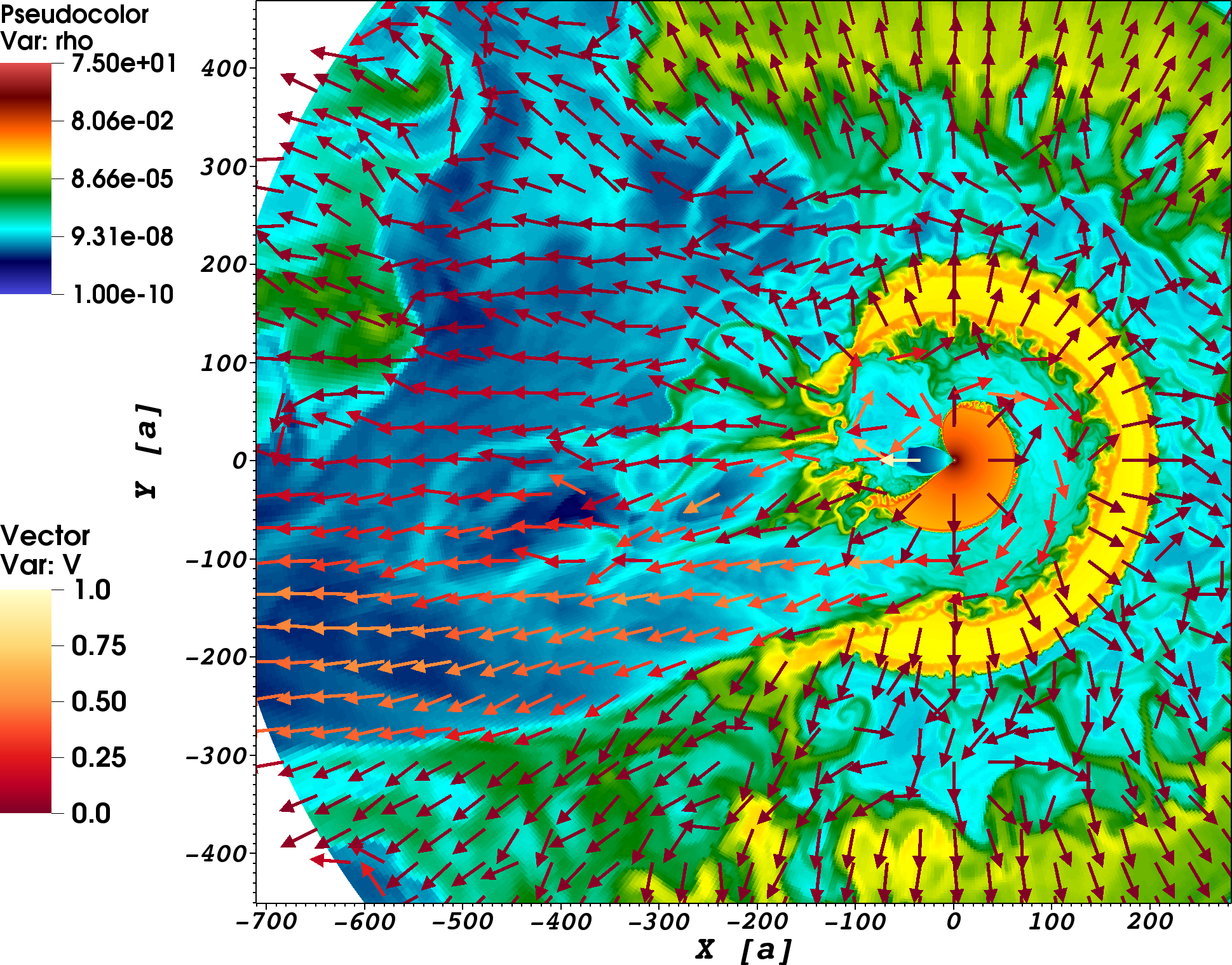}
\caption{The same as in Fig.~\ref{map1} but for $\phi = 0.46$.}
\label{map2}
\end{figure}

Figures~\ref{map1} and \ref{map2} show density maps at periastron {  (orbital phase 0;} right of the grid) and apastron {  (orbital phase 0.5;} left of the grid) for \j06\, computed following the approach 
just presented \citep[see also][]{bbmb17}. {  To obtain these maps, two orbits were simulated to allow the interacting flows to reach a quasi-steady\footnote{Instability growth produces orbit-to-orbit variations.} state.}
There is a shock related to Coriolis forces at the onset of the spiral. This shock is very close to the binary around 
periastron, where the shocked pulsar wind bends, and is barely visible at the very centre of  Fig.~\ref{map1}. One sees in Fig.~\ref{map2}, slightly before apastron, 
that the scale of this shock is much larger, and the pulsar wind is also disrupting the spiral arm in the periastron-apastron direction.

\begin{figure}
\includegraphics[width=0.9\linewidth]{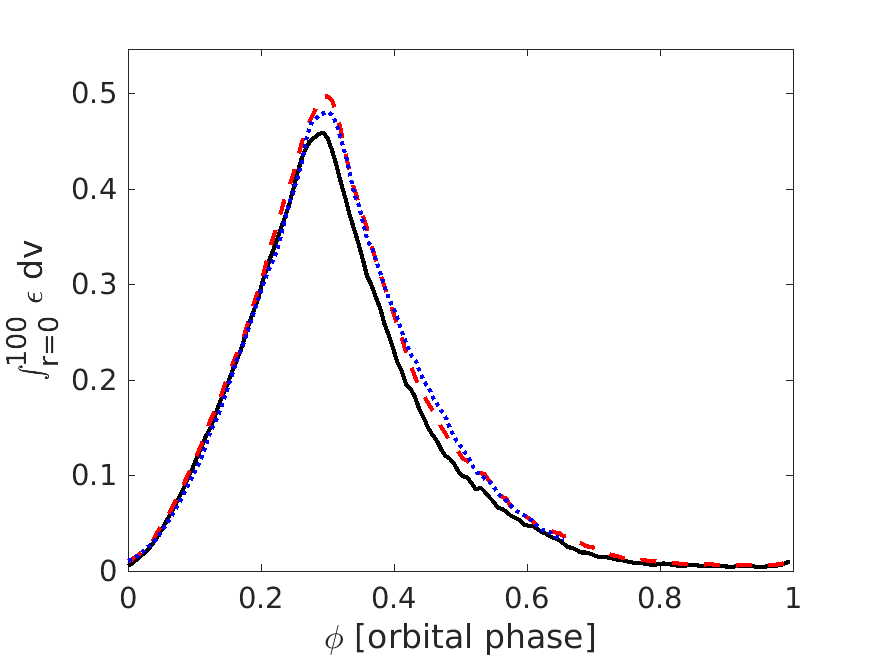}
\caption{Pulsar wind energy accumulated in the first spiral turn for \j06{}.}
\label{lc2}
\end{figure}

Figure~\ref{lc2} \citep[similar to fig.~3 in][]{bbmb17} gives the amount of pulsar wind energy accumulated along the orbit within $100\,a$ (roughly the fist spiral turn), 
whose shape is explained by the growth of the shocked flow, one-arm spiral after periastron, and by the sudden disruption of the spiral arm in the periastron-apastron 
direction around apastron.  As described in the next section, Fig.~\ref{lc2} is in fact giving us information on the non-thermal particles present in the flow assuming 
that they contain a fixed fraction of the total internal energy of the system. Note that the spiral arm evolution along the orbit is characterized by the shocked dense 
stellar wind, which is slow enough to consider that the shocked fast pulsar wind is stationary at a given orbital phase. 

If non-thermal particles are mostly accelerated after the pulsar wind is affected by the spiral turn, the relatively wide nature of the binary, and the shock location, 
allow these particles to flow adiabatically with the shocked pulsar wind but around periastron, when this shock is the closest to the binary, and radiation losses are 
the strongest. Therefore, Fig.~\ref{lc2} can be giving us an accurate description of the non-thermal particle energetics for most of the orbit, and the flow internal 
energy evolution can inform calculations of the non-thermal emission from the shocked pulsar wind structure. Around periastron, some corrections should be introduced 
due to fast radiation cooling.
 
\begin{figure*}
\includegraphics[width=0.47\linewidth]{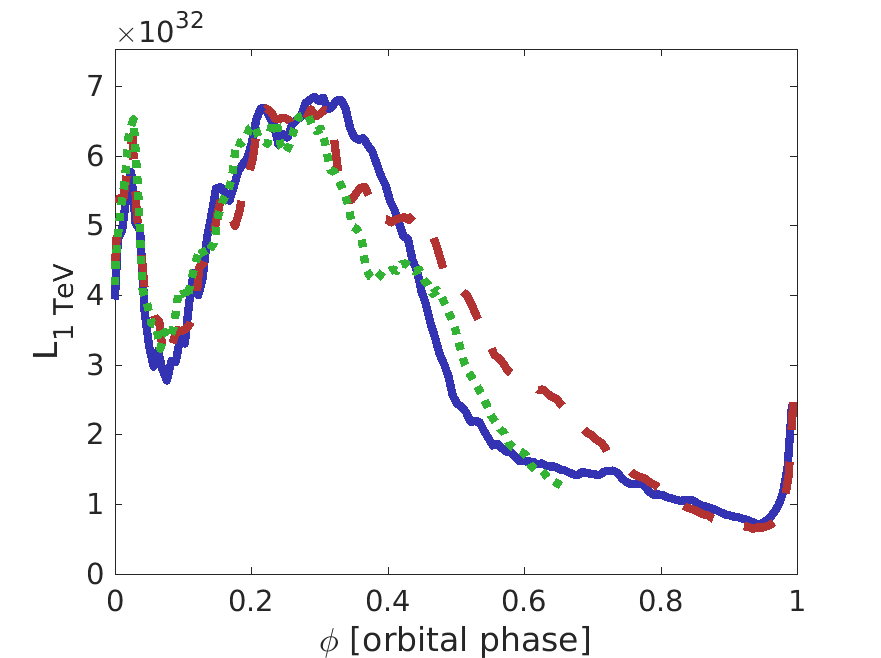}
\includegraphics[width=0.47\linewidth]{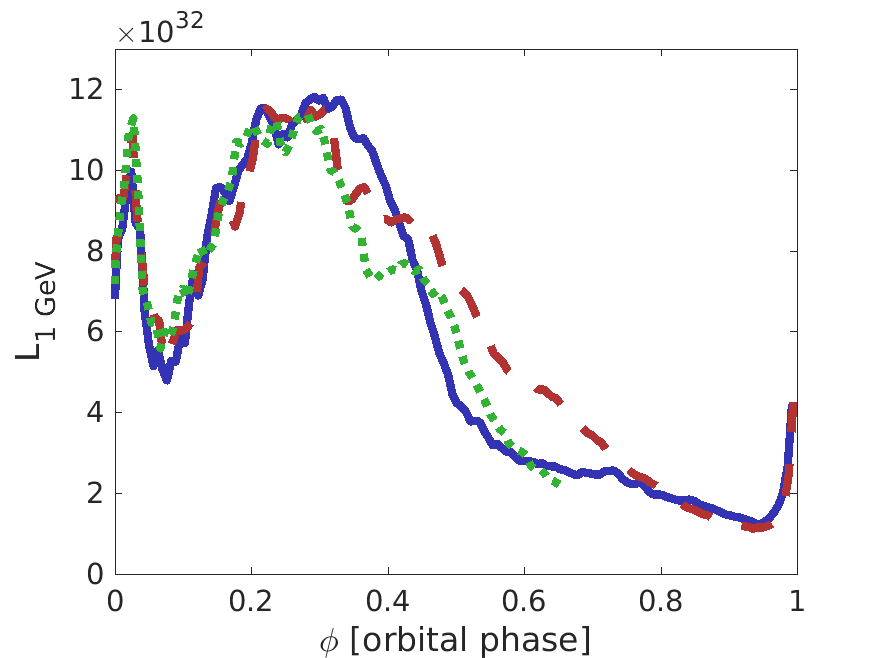}
\includegraphics[width=0.47\linewidth]{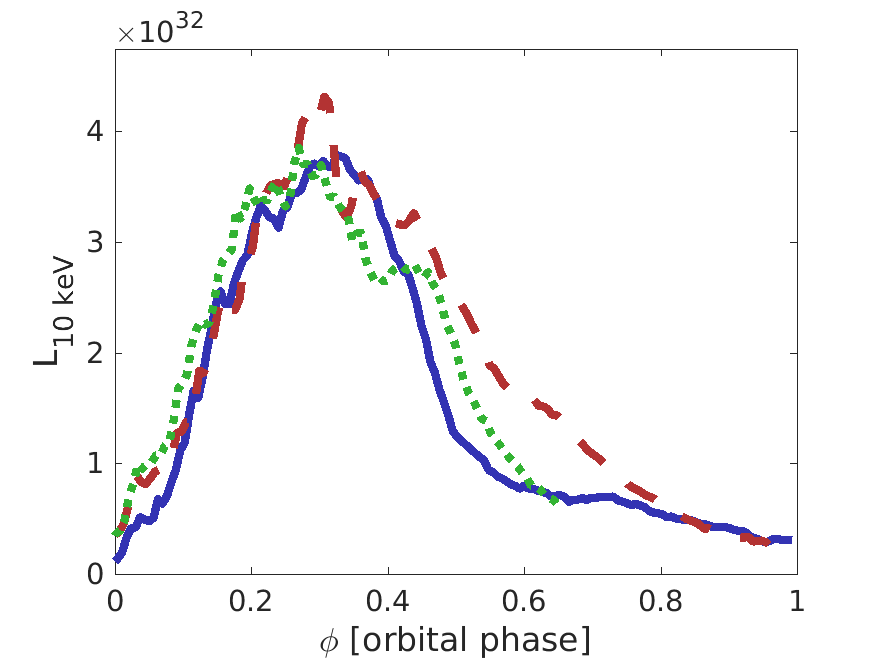}
\includegraphics[width=0.47\linewidth]{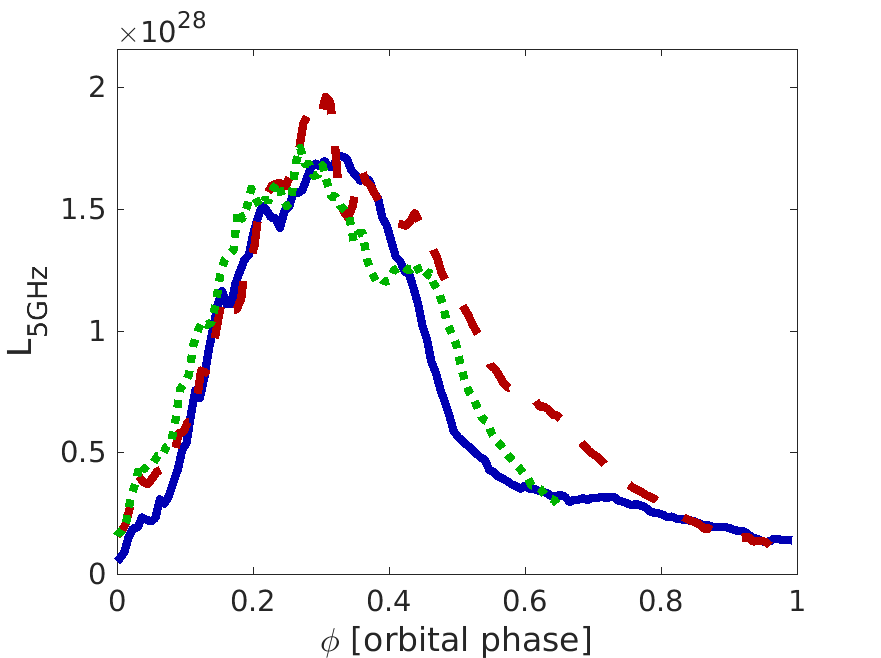}
\caption{Lightcurves for \j06{} at 1~TeV (left top panel), 1~GeV (right top panel), 10~keV (left bottom panel), and 5 GHz (right bottom panel), in cgs units, for three 
contiguous orbits {  (orbit 1: solid line, orbit 2: dashed line; orbit 3: dotted line)}, and $\epsilon_B = 0.5$.}
\label{lc3}
\end{figure*}

\begin{figure*}
\includegraphics[width=0.45\linewidth]{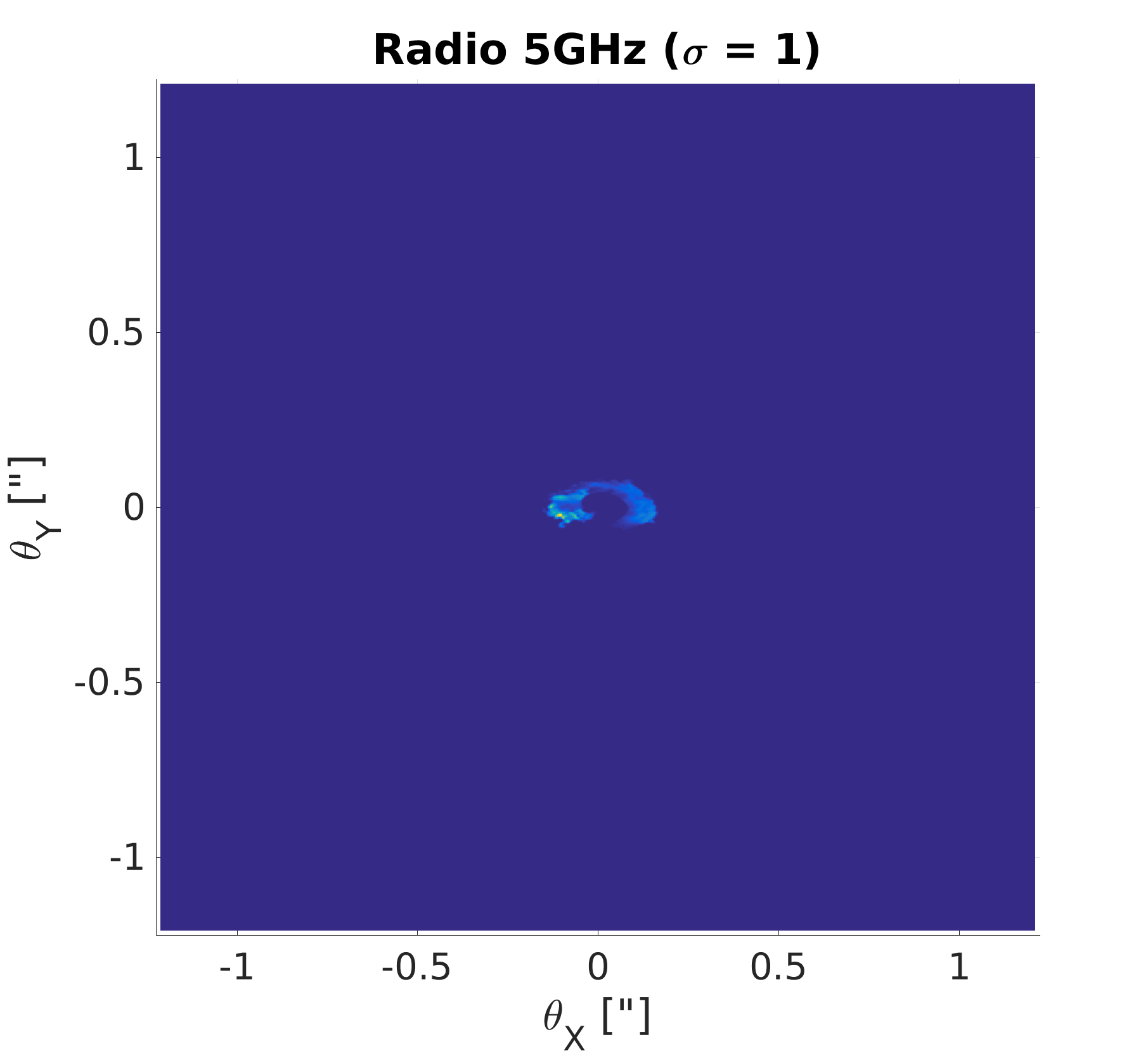}
\includegraphics[width=0.45\linewidth]{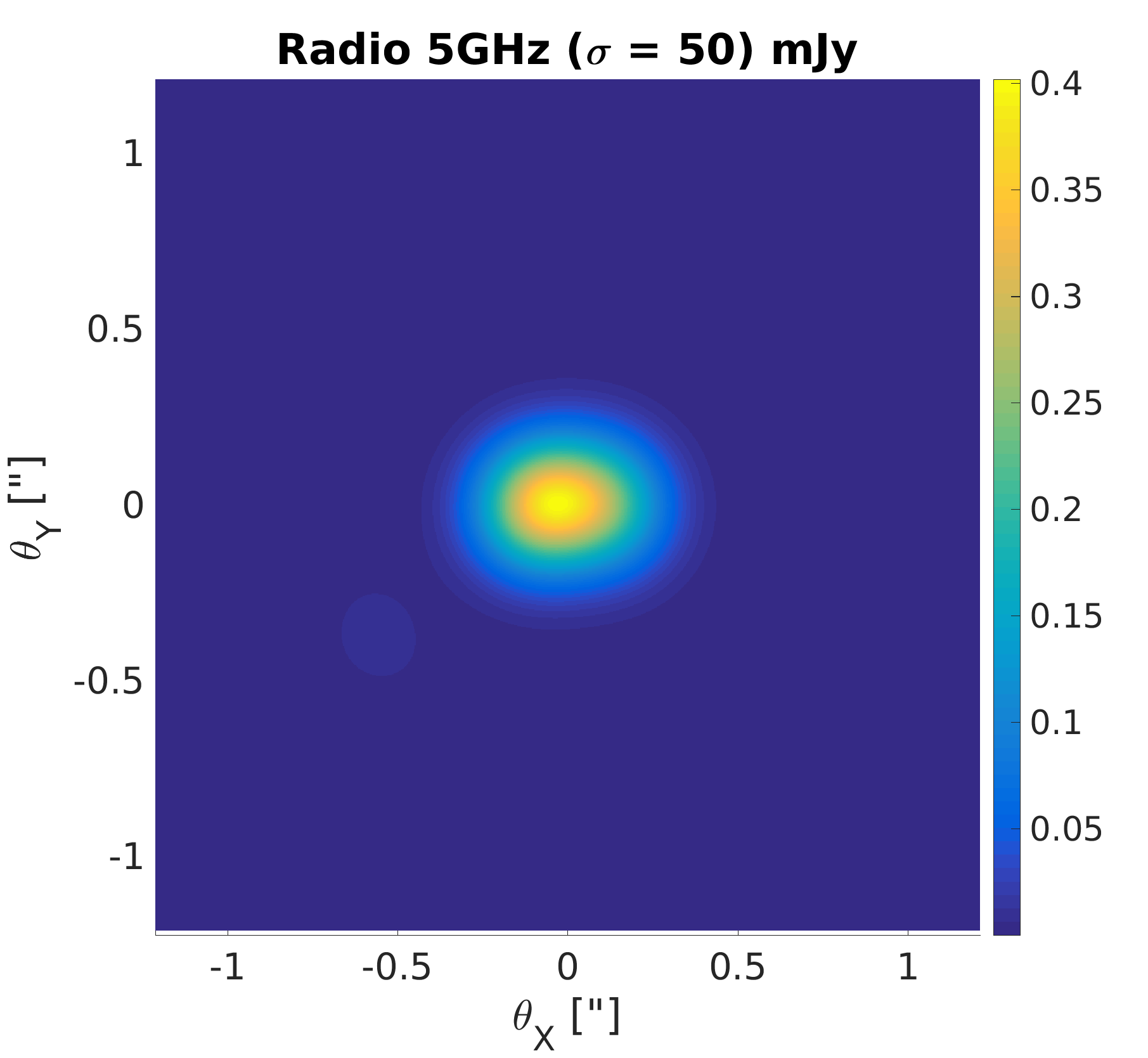}
\includegraphics[width=0.45\linewidth]{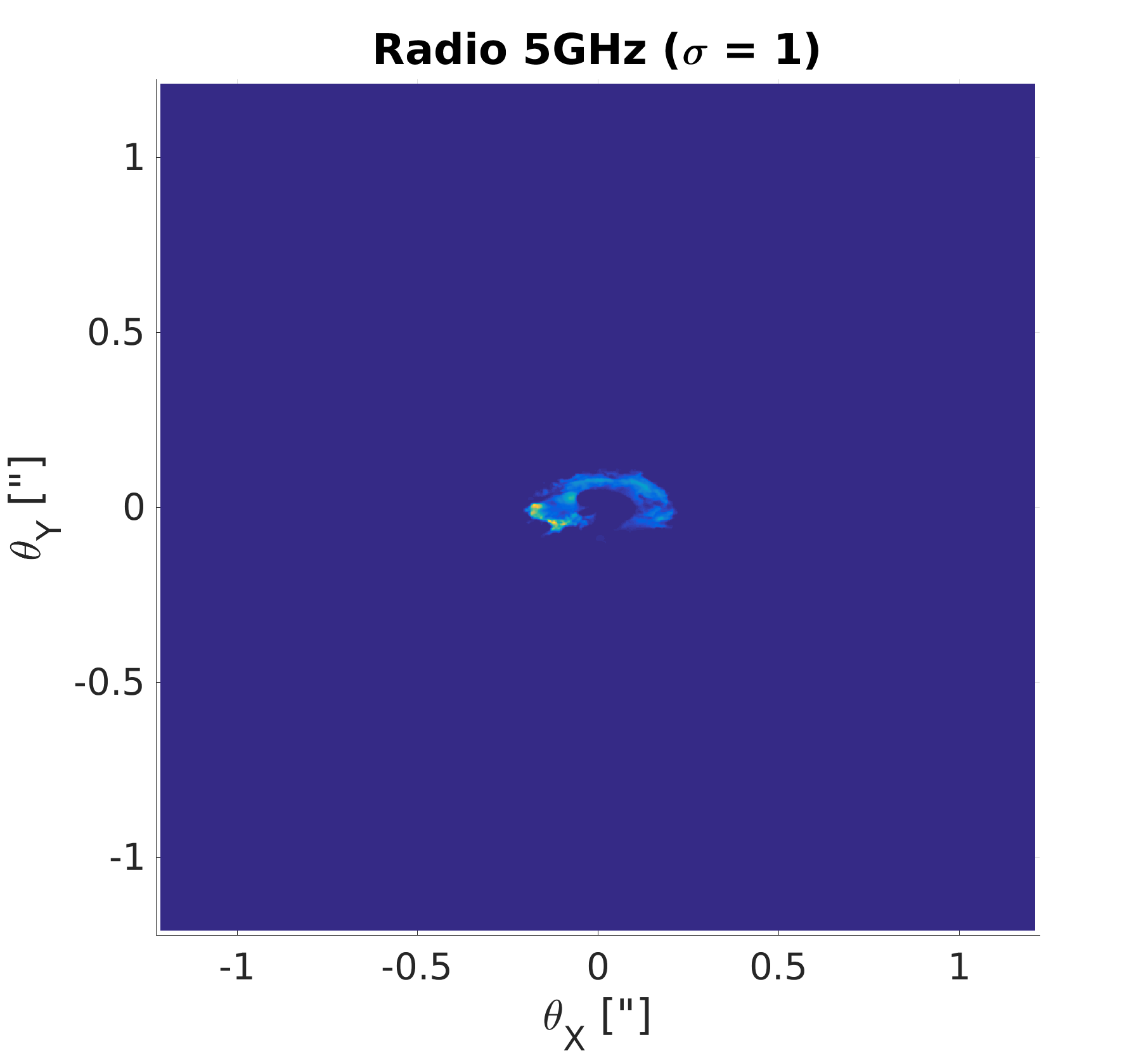}
\includegraphics[width=0.45\linewidth]{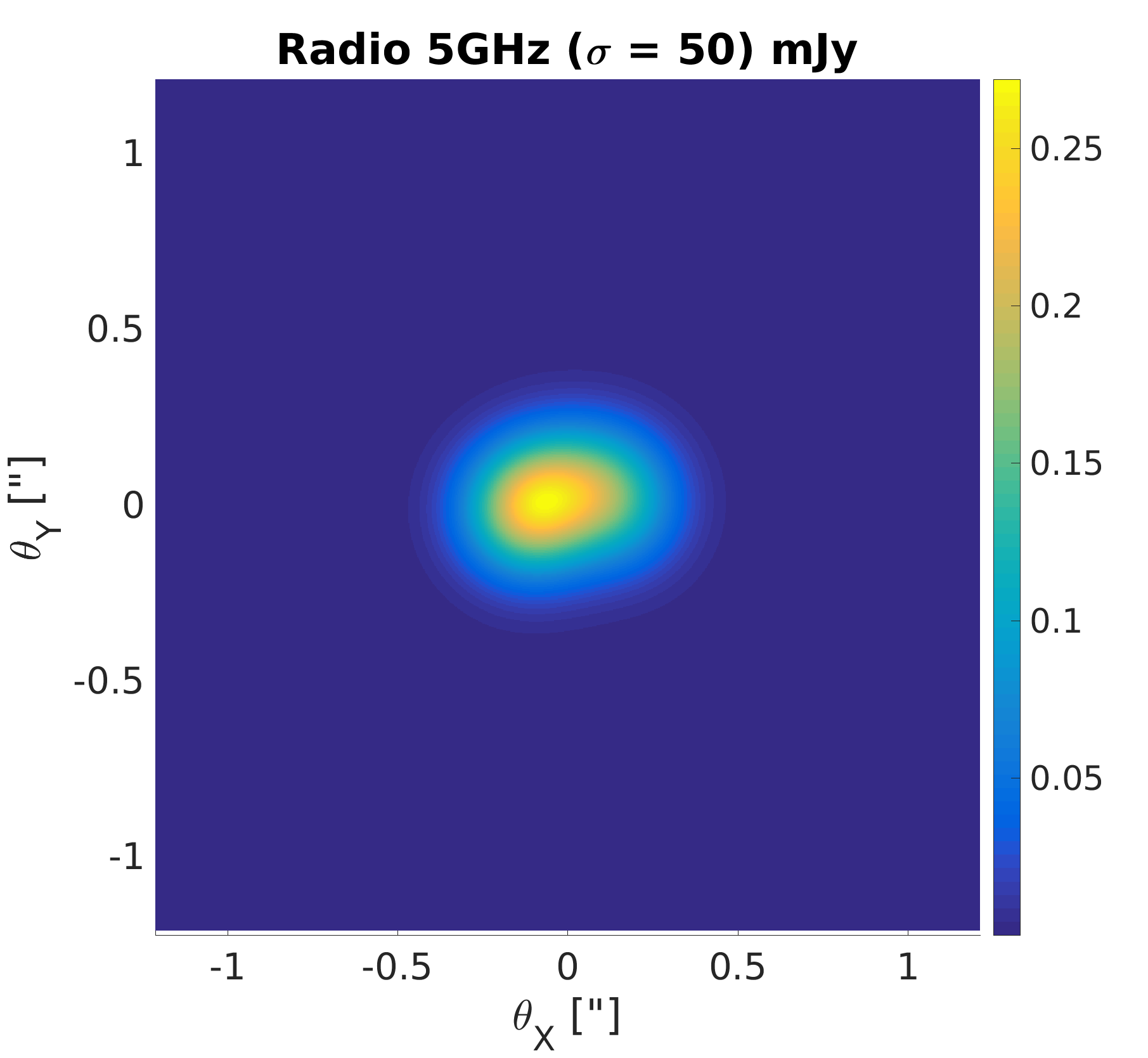}
\includegraphics[width=0.45\linewidth]{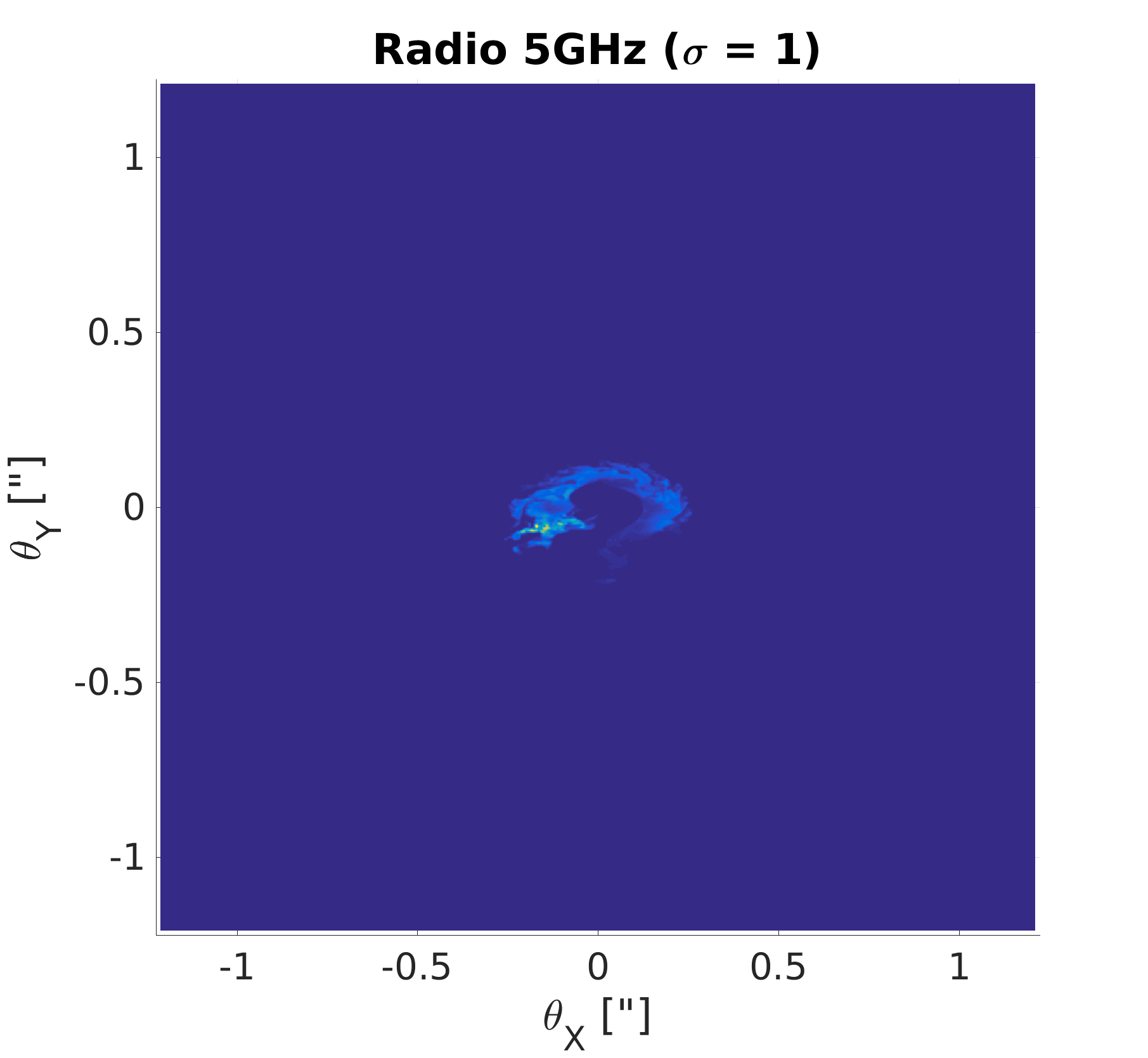}
\includegraphics[width=0.45\linewidth]{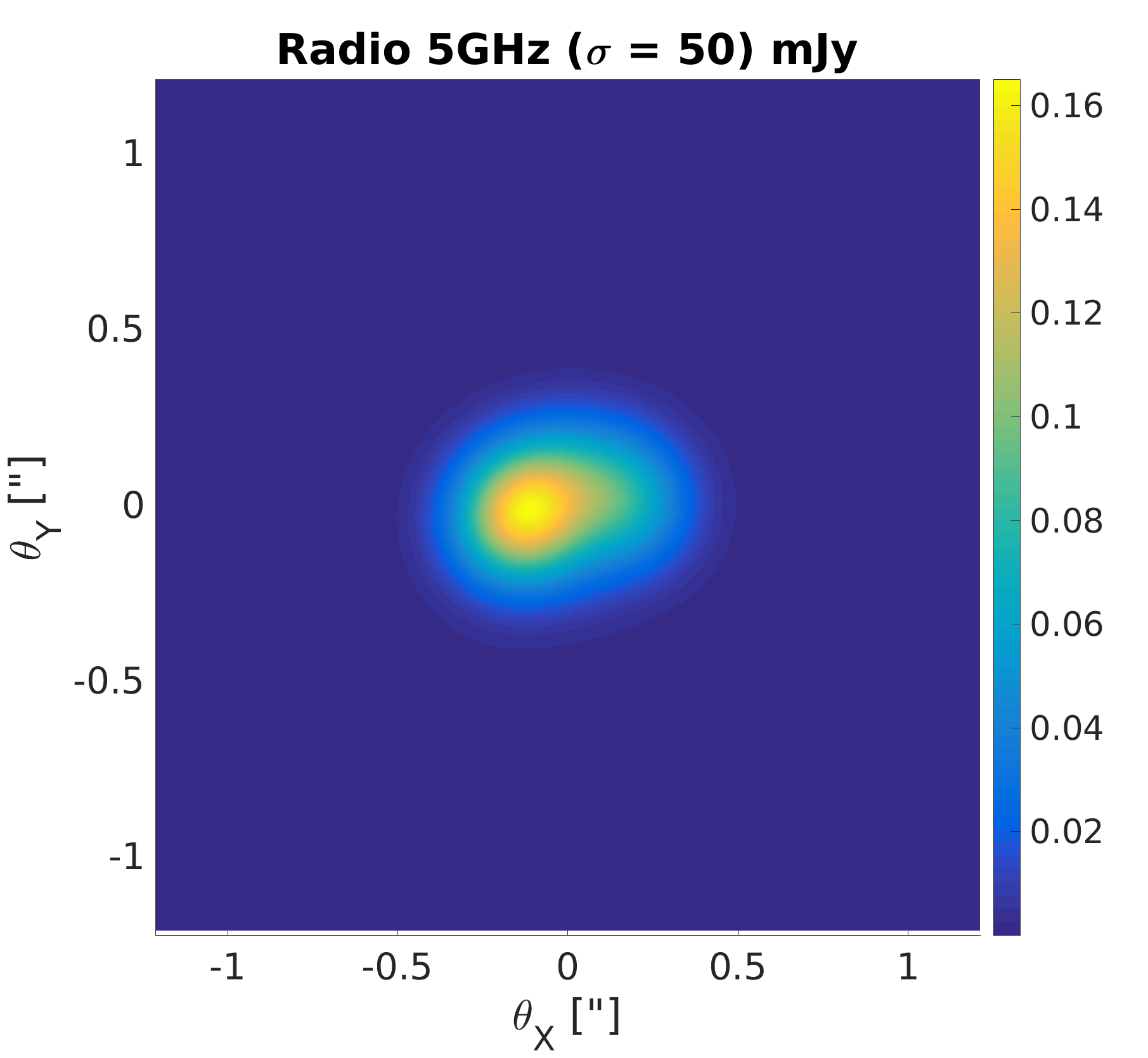}
\caption{Synthetic radio maps with Gaussian filtering $\sigma = 1\,a$ ($\sim 2$~mas; left) and $\sigma = 50\,a$ ($\sim 80$~mas; right), 
at phases 0.3 (top), 0.41 (middle), and 0.52 (bottom); the legend units are in mJy, and $\epsilon_B = 0.5$.}
\label{radmap}
\end{figure*}

\section{Radiation}\label{rad}

The hydrodynamical approach presented above can provide us with semi-quantitative clues of the non-thermal behaviour. The soundness of these clues relies on the 
fact that the equatorial region is energetically dominant, on the assumption that particles are (mostly) accelerated at the {\it Coriolis shock}, and on the validity 
of the adiabatic approximation for the non-thermal energy evolution for most of the orbit. When radiation losses dominate over adiabatic losses near periastron, some 
correction has to be applied. We remark that there are important uncertainties that would render a more accurate approach still very approximate, like inaccurate orbital 
parameters, a not well constrained physics of stellar and the pulsar wind physics, unknown dominant acceleration mechanism and location, and unclear magnetic field conditions. 
In this context, simple analytical prescriptions for the non-thermal processes should suffice at this stage to provide robust trends of the non-thermal emission behaviour.

Here, it is assumed that non-thermal particles are present in the shocked pulsar wind only beyond the point at which the {\it Coriolis} shock bends the flow. Beyond that 
distance, {  the non-thermal energy density of the shocked pulsar wind in a given cell, $u_{\rm cell,NT}$, is first computed in the laboratory frame in the adiabatic approximation:
\be
u_{\rm cell,NT} = \eta_{\rm NT}\,\chi_{\rm pw}\,3P\,\Gamma\,, 
\label{eq:uNT}
\ee
where $3P$ is the (relativistic) internal energy density,} 
$\Gamma$ the bulk Lorentz factor, $\chi_{\rm pw}$ the mass fraction 
of pulsar wind in the cell, and $\eta_{\rm NT}$ an unknown non-thermal particle fraction parameter. Other regions, like the wind-colliding region within the binary, 
may contribute to the lightcurve, but we focus here on the scales of the simulation.

We assume that the non-thermal particles are electrons(/positrons) in the shocked pulsar wind, and focus on synchrotron and inverse Compton (IC) as the relevant emission 
processes at specific particle energies, i.e. those yielding emission at a particular frequency. The corresponding luminosities can be estimated by taking the non-thermal 
energy per cell, and divide it by the corresponding cooling timescale of a particular radiation process at the energy of interest.

For electrons interacting through IC with stellar photons of $\sim 3\,kT\sim 10$~eV (synchrotron photons are negligible as IC targets), the following expression for 
$t_{\rm IC}=\gamma/\dot{\gamma}_{\rm IC}$ is used (adapted from \citealt{bos09}; see also \citealt{kha14}):
$$
\dot{\gamma}_{\rm IC}= 5.5\times10^{17} T_{\rm mcc}^3\gamma \log_{10}(1+0.55\gamma T_{\rm mcc}) \times $$
\be
\qquad \qquad \qquad \frac{1+\frac{1.4 \gamma T_{\rm mcc}}{(1+12 \gamma^2 T_{\rm mcc}^2)}}{1+25\gamma T_{\rm mcc}}\left(\frac{R_*}{r}\right)^2
\label{eq:gicdot}
\ee
where $\gamma$ is the electron Lorentz factor, $T_{\rm mcc} = kT_*/m_ec^2$ the stellar temperature in electron rest energy units, $R_*$ the stellar radius, and $r$ 
the distance to the star, which yields the typical size of the emitting region. This radiation timescale is valid assuming that the target photon field is isotropic or, 
loosely, the angle between the observer and the target direction is not small. As the system inclination is not well known, and the orbital elements may not be fully 
settled, at this stage we neglect angular IC effects.

For synchrotron radiation, the adopted timescale is:
\be
t_{\rm sync}\approx \frac{6\times 10^2}{B^2\,E_{\rm sync}}\,{\rm s},
\label{eq:tsyn}
\ee
where $B$ is the magnetic field in G, assumed to be disordered for simplicity, and
\be
E_{\rm sync}\approx 0.2\left( \frac{\epsilon_{\gamma,keV}}{B\,[G]} \right)^{1/2}\,{\rm erg}.
\label{eq:esy}
\ee
The value of $B$ in the shocked pulsar wind is taken as $B=\sqrt{\epsilon_B 8 \pi P}$, where $\epsilon_B\le 1$ is a non-dimensional normalization parameter. 
Values $\epsilon_B\sim 1$ would render the hydrodynamics assumption invalid.

The IC and synchrotron lightcurves can be calculated integrating the emissivity of each cell, $u_{\rm cell,NT}/t_{\rm IC/sync}$, over the emitting volume: 
$L_{\rm IC/sync}\sim\int_V(u_{\rm cell,NT}/t_{\rm IC/sync}) dV$, {  where $dV$ is the cell volume in the laboratory frame}. This is approximately correct for an electron injection distribution in energy $\propto 
E^{-2}$, which is common for non-thermal astrophysical sources. Much harder (softer) energy distributions would enhance the synchrotron and the IC fluxes at the 
highest (lowest) energy bands. {  We note that the electron energy distribution inferred by \cite{mal17} through model fitting of multiwavelength data goes from $\propto E^{-1.3}$ at low energies to $\propto E^{-2.3}$ at high energies, although the observed X-ray photon index presents significant changes along the orbit around $\sim 1.5$ (see fig.~3 in that work), i.e. $\propto E^{-2}$. An accurate treatment of the electron energy distribution is important, but it is left for future studies.} Given that the bulk velocity of the shocked pulsar wind is mildly relativistic at most, the semi-quantitative nature of our approach, and 
the uncertainties in the binary and the binary-observer geometry, we neglect at this stage Doppler boosting effects.

The expressions given for $L_{\rm IC/sync}$ are valid under dominant adiabatic losses. For a certain region of size $\sim r$, one can estimate the non-radiative losses 
(adiabatic losses and escape) as
\be
t_{\rm nonrad} \sim r / \vv_{\rm f}\,,
\label{eq:tnonrad}
\ee
where $\vv_{\rm f}$ is the typical flow dynamical velocity, which is
\be
\vv_{\rm f} = \max(\vv_{\rm r},0.1\,\vv_{\rm t})\,,
\label{eq:vf}
\ee
with $\vv_{\rm r}$ being the radial velocity, and $\vv_{\rm t}\approx\sqrt{\vv_{\rm r}^2 + \vv_{\theta}^2 + \vv_{\phi}^2}$. This prescription is adopted to account 
for expansion of the spiral arms even in regions in which $v_{\rm r}$ is small or negative. 

If $t_{\rm nonrad}>t_{\rm rad}$, where  $t_{\rm rad}=1/(1/t_{\rm IC}+1/t_{\rm sync})$, the adiabatic approximation breaks. In the cells where it happens, $u_{\rm NT}$ 
is depleted by radiation losses, and only a fraction
$f\sim \min(1,t_{\rm rad}/t_{\rm nonrad})$ of the non-thermal energy would be present. To account for this, $f$ is incorporated to the luminosity calculation as:  
\be
L_{\rm IC/sync} \sim a_{\rm Band} \int_V \frac{ f\, u_{\rm cell,NT}}{t_{\rm IC/sync}} \,dV\,.
\label{eq:LICn}
\ee
an additional factor $a_{\rm Band}\sim 0.1$ is also included to account for the narrow range in energy considered of the broadband radiation. 

We assume that the non-thermal particles are injected at twice the turnover radius $x$ \citep{bb11,bbkp12}, roughly the location at which the shocked pulsar wind strongly 
bends, where $x= (3/2)\, \vv_{\rm w} \eta_{\rm PW}^{1/2}/\Omega,$, with $\vv_{\rm w}$ being the stellar wind velocity, $\eta_{\rm PW} = 0.1$ the pulsar-to-stellar wind 
thrust ratio, and $\Omega$ the orbital angular velocity \citep[see][]{bbp15}. Thus, we integrate the volume in Eq.~(\ref{eq:LICn}) from $2\,x$ up to the outer computational 
boundary. Note that even when particles of high enough energy quickly cool down close to the binary around periastron, $f$ does not reduce the emission from farther distances, 
as it should if particles are accelerated around $2\,x$, and therefore our calculation overestimates $L_{\rm IC/sync}$ around periastron. 

In this work we focus in the energy bands around 5~GHz and 10~keV for the synchrotron emission, and 1~GeV and 1~TeV for IC. The lightcurves of three contiguous simulated orbits are presented in Fig.~\ref{lc3}. {  The three lightcurves are similar, which means that they are not much sensitive to the initial conditions of the first orbit, nor to the orbit-by-orbit variations of the spiral structure.} Figure~\ref{radmap} shows the radio maps at 5~GHz for three different orbital phases: 0.3, 0.41, and 0.52, and two different Gaussian filters that mimic a radio interferometric beam: $\sigma = 1\,a$ ($\sim 2$~mas) and $\sigma = 50\,a$ ($\sim 80$~mas). {  In both Figs~\ref{lc3} and \ref{radmap}, $\epsilon_B$ and $\eta_{\rm NT}$ have been fixed to 0.5 and 0.1, respectively, keeping them constant with time and in the whole computational domain.}

\section{Summary and discussion}\label{disc}

Using hydrodynamical information, we have computed the non-thermal emission in radio, X-rays, and gamma rays for the gamma-ray binary \j06, in the pulsar scenario, and 
focusing on particles accelerated at the spiral onset. Remarkably, the multiwavelength predicted lightcurves and fluxes, and even morphology (radio), are in rather good 
agreement with observations \citep[e.g.][]{2011A&A...533L...7M,ali14,mal17,mai17}, in particular taking into account the approximate nature of our radiation prescriptions.

All the presented lightcurves show a peak around phases 0.3--0.4, which is connected with
the accumulation of the hot shocked plasma in the inner spiral arm, and its later  quick
release as the spiral arm is disrupted in the periastron-apastron direction. In the IC
lightcurves (GeV and TeV) a narrow peak is also present right after periastron passage,
possibly because of high IC efficiency at $r\gtrsim 2\,x$. However, the IC peak may be an
artifact of our simple way of including radiation cooling, but we cannot completely 
discard its presence without more realistic radiation calculations. Nevertheless, for the
way how  radiation cooling is treated here (i.e. the $f$ factor in Eq.~\ref{eq:LICn}),  we
are quite confident that the emission around periastron is overestimated rather than
underestimated. It is worth noting however that a weak hint of a small TeV peak around 
phases $\sim 0.0-0.1$ is seen in \cite{mai17}, when folding in phase MAGIC data from
\cite{ale12} with a newer orbital period. {  The computed lightcurves do not reproduce the broad secondary peak in the X-ray and the TeV lightcurves \citep[e.g.][]{ali14}. We recall nevertheless that we do not consider the physics at the scales of the binary, on which the Be disc, among other factors, could play a relevant role. In addition, our simplified approach to hydrodynamics, and its coupling with radiation, may be hiding effects that would arise in more complicated calculations.}

The radio maps unveil an extended structure moving in the periastron-apastron direction. The radio flux evolution, and the displacement 
of the radio structure on the studied spatial scales, are consistent with those found by \cite{2011A&A...533L...7M}. 
It is predicted that the motion of the observed radio structure should be aligned with the periastron-apastron direction. Moreover, a weak ellipsoidal  
extension is hinted  also in the apastron-periastron direction, formed by the half ring of shocked pulsar wind present at the periastron side seen in 
Figs.~\ref{map1} and \ref{map2}. Therefore,  X-ray \citep{bb16} and radio \citep[see also][]{bbmb17} moving structures may appear prominently at 
the apastron side, with a fainter counterpart at the periastron side.

{  The scenario presented here would not apply if the binary were less eccentric, as proposed that \j06\, may be by \cite{mor18}, and discussed also in 
\cite{mal17} in the context of X-ray data. In a less eccentric binary (our calculations point at $e\lesssim 0.75$), the first spiral arm at the apastron 
side would survive apastron, and a strong lightcurve peak would not be expected before that orbital phase. In principle, the qualitative agreement of our 
predictions with observations may suggest that the system is indeed very eccentric and hosts a young pulsar, although the uncertainty in the orbital 
elements has to be reduced to confirm the high eccentricity. Such a high eccentricity would, in the scenario presented here, strengthen the case for 
a pulsar in the system.
 
\cite{mal17} and \cite{mor18} considered the possibility of an inclined Be disc playing an important role in the high-energy emission. 
If the Be disc mass beyond the pulsar location were not much smaller than that injected by the stellar wind during $T$, then we would tentatively expect 
an even stronger accumulation of the pulsar wind energy before apastron because of the disc mass, disrupting the spiral faster. This could make the 
lightcurve peak before apastron somewhat higher, narrower, and earlier. Nevertheless, the role of a rather massive disc, and its inclination, can only 
be properly assessed through numerical calculations. Interestingly, \cite{mor18} and \cite{mal17} also proposed that the periastron and apastron orbital 
phases are significantly different from those in \cite{cas12}. If this were the case, then the mechanism proposed by us could be hardly behind the dominant 
lightcurve peak of \j06, regardless of the eccentricity.
In the particular case the periastron were $\sim 0.3$ in phase earlier, the narrow GeV and TeV predicted peaks may actually be the first GeV and TeV peaks 
of the observed lightcurve, and the broader predicted X-ray and TeV peaks could correspond to the second observed X-ray and TeV peaks of the lightcurve. In that
case, the first X-ray peak of the lightcurve may need adding an extra emitting components (e.g. the binary region, Be disc interaction, etc.). 
In this scenario, the radio behaviour would be more difficult to explain though.
}

A next step would be to carry out more accurate radiation calculations applying
post-processing to the hydrodynamical solution for \j06, and compute consistently the
particle evolution in space and energy in a way similar to those employed for instance by
\cite{dub15} and \cite{del17}. Another step forward would be to increase the resolution in
the $\theta$-coordinate when simulating the wind collision at large scales in this
source. {  Note that including the Be disc in the simulations would require including the binary region and a Cartesian grid, making simulations much more demanding.}

\section{Acknowledgments}
We acknowledge the help of Dmitry Khangulyan with the optimization of our post-processing procedure and useful discussions. 
This work was supported by NSF grant AST-1306672, 
DoE grant DE-SC0016369, and NASA grant 80NSSC17K0757 and GO8-19040X.
This work was also supported by the Spanish Ministerio de Econom\'{i}a y Competitividad (MINECO/FEDER, UE) 
under grants AYA2013-47447-C3-1-P and AYA2016-76012-C3-1-P, 
MDM-2014-0369 of ICCUB (Unidad de Excelencia `Mar\'{i}a de Maeztu'), and the Catalan DEC grant 2014 SGR 86. 

\bibliographystyle{mnras}
\bibliography{text}   %expects file references.bib

\begin{thebibliography}{}
\makeatletter
\relax
\def\mn@urlcharsother{\let\do\@makeother \do\$\do\&\do\#\do\^\do\_\do\%\do\~}
\def\mn@doi{\begingroup\mn@urlcharsother \@ifnextchar [ {\mn@doi@}
  {\mn@doi@[]}}
\def\mn@doi@[#1]#2{\def\@tempa{#1}\ifx\@tempa\@empty \href
  {http://dx.doi.org/#2} {doi:#2}\else \href {http://dx.doi.org/#2} {#1}\fi
  \endgroup}
\def\mn@eprint#1#2{\mn@eprint@#1:#2::\@nil}
\def\mn@eprint@arXiv#1{\href {http://arxiv.org/abs/#1} {{\tt arXiv:#1}}}
\def\mn@eprint@dblp#1{\href {http://dblp.uni-trier.de/rec/bibtex/#1.xml}
  {dblp:#1}}
\def\mn@eprint@#1:#2:#3:#4\@nil{\def\@tempa {#1}\def\@tempb {#2}\def\@tempc
  {#3}\ifx \@tempc \@empty \let \@tempc \@tempb \let \@tempb \@tempa \fi \ifx
  \@tempb \@empty \def\@tempb {arXiv}\fi \@ifundefined
  {mn@eprint@\@tempb}{\@tempb:\@tempc}{\expandafter \expandafter \csname
  mn@eprint@\@tempb\endcsname \expandafter{\@tempc}}}

\bibitem[\protect\citeauthoryear{{Aharonian}, {Akhperjanian}, {Aye}  \&
  {et~al.}}{{Aharonian} et~al.}{2005}]{aha05}
{Aharonian} F.,  {Akhperjanian} A.~G.,  {Aye} K.-M.,   {et~al.} 2005, \mn@doi
  [\aap] {10.1051/0004-6361:20052983}, \href
  {http://adsabs.harvard.edu/abs/2005A%26A...442....1A} {442, 1}

\bibitem[\protect\citeauthoryear{{Aharonian}, {Akhperjanian}, {Bazer-Bachi}  \&
  {et~al.}}{{Aharonian} et~al.}{2007}]{aha07}
{Aharonian} F.~A.,  {Akhperjanian} A.~G.,  {Bazer-Bachi} A.~R.,   {et~al.}
  2007, \mn@doi [\aap] {10.1051/0004-6361:20077299}, \href
  {http://adsabs.harvard.edu/abs/2007A%26A...469L...1A} {469, L1}

\bibitem[\protect\citeauthoryear{{Aleksi{\'c}}, {Alvarez}, {Antonelli}  \&
  {et~al.}}{{Aleksi{\'c}} et~al.}{2012}]{ale12}
{Aleksi{\'c}} J.,  {Alvarez} E.~A.,  {Antonelli} L.~A.,   {et~al.} 2012,
  \mn@doi [\apjl] {10.1088/2041-8205/754/1/L10}, \href
  {http://adsabs.harvard.edu/abs/2012ApJ...754L..10A} {754, L10}

\bibitem[\protect\citeauthoryear{{Aliu}, {Archambault}, {Aune}  \&
  {et~al.}}{{Aliu} et~al.}{2014}]{ali14}
{Aliu} E.,  {Archambault} S.,  {Aune} T.,   {et~al.} 2014, \mn@doi [\apj]
  {10.1088/0004-637X/780/2/168}, \href
  {http://adsabs.harvard.edu/abs/2014ApJ...780..168A} {780, 168}

\bibitem[\protect\citeauthoryear{{Barkov} \& {Bosch-Ramon}}{{Barkov} \&
  {Bosch-Ramon}}{2016}]{bb16}
{Barkov} M.~V.,  {Bosch-Ramon} V.,  2016, \mn@doi [\mnras]
  {10.1093/mnrasl/slv171}, \href
  {http://adsabs.harvard.edu/abs/2016MNRAS.456L..64B} {456, L64}

\bibitem[\protect\citeauthoryear{{Bongiorno}, {Falcone}, {Stroh}  \&
  {et~al.}}{{Bongiorno} et~al.}{2011}]{bon11}
{Bongiorno} S.~D.,  {Falcone} A.~D.,  {Stroh} M.,   {et~al.} 2011, \mn@doi
  [\apjl] {10.1088/2041-8205/737/1/L11}, \href
  {http://adsabs.harvard.edu/abs/2011ApJ...737L..11B} {737, L11}

\bibitem[\protect\citeauthoryear{{Bosch-Ramon} \& {Barkov}}{{Bosch-Ramon} \&
  {Barkov}}{2011}]{bb11}
{Bosch-Ramon} V.,  {Barkov} M.~V.,  2011, \mn@doi [\aap]
  {10.1051/0004-6361/201117235}, \href
  {http://adsabs.harvard.edu/abs/2011A%26A...535A..20B} {535, A20}

\bibitem[\protect\citeauthoryear{{Bosch-Ramon} \& {Khangulyan}}{{Bosch-Ramon}
  \& {Khangulyan}}{2009}]{bos09}
{Bosch-Ramon} V.,  {Khangulyan} D.,  2009, \mn@doi [International Journal of
  Modern Physics D] {10.1142/S0218271809014601}, \href
  {http://adsabs.harvard.edu/abs/2009IJMPD..18..347B} {18, 347}

\bibitem[\protect\citeauthoryear{{Bosch-Ramon}, {Barkov}, {Khangulyan}  \&
  {Perucho}}{{Bosch-Ramon} et~al.}{2012}]{bbkp12}
{Bosch-Ramon} V.,  {Barkov} M.~V.,  {Khangulyan} D.,   {Perucho} M.,  2012,
  \mn@doi [\aap] {10.1051/0004-6361/201219251}, \href
  {http://adsabs.harvard.edu/abs/2012A%26A...544A..59B} {544, A59}

\bibitem[\protect\citeauthoryear{{Bosch-Ramon}, {Barkov}  \&
  {Perucho}}{{Bosch-Ramon} et~al.}{2015}]{bbp15}
{Bosch-Ramon} V.,  {Barkov} M.~V.,   {Perucho} M.,  2015, \mn@doi [\aap]
  {10.1051/0004-6361/201425228}, \href
  {http://adsabs.harvard.edu/abs/2015A%26A...577A..89B} {577, A89}

\bibitem[\protect\citeauthoryear{{Bosch-Ramon}, {Barkov}, {Mignone}  \&
  {Bordas}}{{Bosch-Ramon} et~al.}{2017}]{bbmb17}
{Bosch-Ramon} V.,  {Barkov} M.~V.,  {Mignone} A.,   {Bordas} P.,  2017, \mn@doi
  [\mnras] {10.1093/mnrasl/slx124}, \href
  {http://adsabs.harvard.edu/abs/2017MNRAS.471L.150B} {471, L150}

\bibitem[\protect\citeauthoryear{{Casares}, {Rib{\'o}}, {Ribas}, {Paredes},
  {Vilardell}  \& {Negueruela}}{{Casares} et~al.}{2012}]{cas12}
{Casares} J.,  {Rib{\'o}} M.,  {Ribas} I.,  {Paredes} J.~M.,  {Vilardell} F.,
  {Negueruela} I.,  2012, \mn@doi [\mnras] {10.1111/j.1365-2966.2011.20368.x},
  \href {http://adsabs.harvard.edu/abs/2012MNRAS.421.1103C} {421, 1103}

\bibitem[\protect\citeauthoryear{{Dubus}, {Lamberts}  \& {Fromang}}{{Dubus}
  et~al.}{2015}]{dub15}
{Dubus} G.,  {Lamberts} A.,   {Fromang} S.,  2015, \mn@doi [\aap]
  {10.1051/0004-6361/201425394}, \href
  {http://adsabs.harvard.edu/abs/2015A%26A...581A..27D} {581, A27}

\bibitem[\protect\citeauthoryear{{Hinton}, {Skilton}, {Funk}  \&
  {et~al.}}{{Hinton} et~al.}{2009}]{hin09}
{Hinton} J.~A.,  {Skilton} J.~L.,  {Funk} S.,   {et~al.} 2009, \mn@doi [\apjl]
  {10.1088/0004-637X/690/2/L101}, \href
  {http://adsabs.harvard.edu/abs/2009ApJ...690L.101H} {690, L101}

\bibitem[\protect\citeauthoryear{{Khangulyan}, {Aharonian}  \&
  {Kelner}}{{Khangulyan} et~al.}{2014}]{kha14}
{Khangulyan} D.,  {Aharonian} F.~A.,   {Kelner} S.~R.,  2014, \mn@doi [\apj]
  {10.1088/0004-637X/783/2/100}, \href
  {http://adsabs.harvard.edu/abs/2014ApJ...783..100K} {783, 100}

\bibitem[\protect\citeauthoryear{{Li}, {Torres}, {Cheng}, {de Ona Wilhelmi},
  {Kretschmar}, {Hou}  \& {Takata}}{{Li} et~al.}{2017}]{li17}
{Li} J.,  {Torres} D.~F.,  {Cheng} K.-S.,  {de Ona Wilhelmi} E.,  {Kretschmar}
  P.,  {Hou} X.,   {Takata} J.,  2017, preprint, \href
  {http://adsabs.harvard.edu/abs/2017arXiv170704280L} {} (\mn@eprint {arXiv}
  {1707.04280})

\bibitem[\protect\citeauthoryear{{Maier} \& {the VERITAS
  Collaboration}}{{Maier} \& {the VERITAS Collaboration}}{2017}]{mai17}
{Maier} G.,  {the VERITAS Collaboration} 2017, preprint, \href
  {http://adsabs.harvard.edu/abs/2017arXiv170804045M} {} (\mn@eprint {arXiv}
  {1708.04045})

\bibitem[\protect\citeauthoryear{{Malyshev} \& {Chernyakova}}{{Malyshev} \&
  {Chernyakova}}{2016}]{mal16}
{Malyshev} D.,  {Chernyakova} M.,  2016, \mn@doi [\mnras]
  {10.1093/mnras/stw2173}, \href
  {http://adsabs.harvard.edu/abs/2016MNRAS.463.3074M} {463, 3074}

\bibitem[\protect\citeauthoryear{{Malyshev}, {Chernyakova}, {Santangelo}  \&
  {P{\"u}hlhofer}}{{Malyshev} et~al.}{2017}]{mal17}
{Malyshev} D.,  {Chernyakova} M.,  {Santangelo} A.,   {P{\"u}hlhofer} G.,
  2017, preprint, \href {http://adsabs.harvard.edu/abs/2017arXiv171105001M} {}
  (\mn@eprint {arXiv} {1711.05001})

\bibitem[\protect\citeauthoryear{{Miller-Jones} et~al.,}{{Miller-Jones}
  et~al.}{2018}]{mil18}
{Miller-Jones} J.~C.~A.,  et~al., 2018, preprint, \href
  {http://adsabs.harvard.edu/abs/2018arXiv180408402M} {} (\mn@eprint {arXiv}
  {1804.08402})

\bibitem[\protect\citeauthoryear{{Mirzoyan} \& {Mukherjee}}{{Mirzoyan} \&
  {Mukherjee}}{2017}]{2017ATel10971....1M}
{Mirzoyan} R.,  {Mukherjee} R.,  2017, The Astronomer's Telegram, \href
  {http://adsabs.harvard.edu/abs/2017ATel10971....1M} {10971}

\bibitem[\protect\citeauthoryear{{Mold{\'o}n}, {Rib{\'o}}  \&
  {Paredes}}{{Mold{\'o}n} et~al.}{2011}]{2011A&A...533L...7M}
{Mold{\'o}n} J.,  {Rib{\'o}} M.,   {Paredes} J.~M.,  2011, \mn@doi [\aap]
  {10.1051/0004-6361/201117764}, \href
  {http://adsabs.harvard.edu/abs/2011A%26A...533L...7M} {533, L7}

\bibitem[\protect\citeauthoryear{{Moritani}, {Kawano}, {Chimasu}, {Kawachi},
  {Takahashi}, {Takata}  \& {Carciofi}}{{Moritani} et~al.}{2018}]{mor18}
{Moritani} Y.,  {Kawano} T.,  {Chimasu} S.,  {Kawachi} A.,  {Takahashi} H.,
  {Takata} J.,   {Carciofi} A.~C.,  2018, preprint, \href
  {http://adsabs.harvard.edu/abs/2018arXiv180403831M} {} (\mn@eprint {arXiv}
  {1804.03831})

\bibitem[\protect\citeauthoryear{{Pavlov}, {Hare}, {Kargaltsev}, {Rangelov}  \&
  {Durant}}{{Pavlov} et~al.}{2015}]{pav15}
{Pavlov} G.~G.,  {Hare} J.,  {Kargaltsev} O.,  {Rangelov} B.,   {Durant} M.,
  2015, \mn@doi [\apj] {10.1088/0004-637X/806/2/192}, \href
  {http://adsabs.harvard.edu/abs/2015ApJ...806..192P} {806, 192}

\bibitem[\protect\citeauthoryear{{Yi} \& {Cheng}}{{Yi} \& {Cheng}}{2017}]{yi17}
{Yi} S.-X.,  {Cheng} K.~S.,  2017, \mn@doi [\mnras] {10.1093/mnras/stx1928},
  \href {http://adsabs.harvard.edu/abs/2017MNRAS.471.4228Y} {471, 4228}

\bibitem[\protect\citeauthoryear{{de la Cita}, {Bosch-Ramon},
  {Paredes-Fortuny}, {Khangulyan}  \& {Perucho}}{{de la Cita}
  et~al.}{2017}]{del17}
{de la Cita} V.~M.,  {Bosch-Ramon} V.,  {Paredes-Fortuny} X.,  {Khangulyan} D.,
    {Perucho} M.,  2017, \mn@doi [\aap] {10.1051/0004-6361/201629112}, \href
  {http://adsabs.harvard.edu/abs/2017A%26A...598A..13D} {598, A13}

\makeatother
\end{thebibliography}

\end{document}